\documentclass[final,3p,times,twocolumn,authoryear]{elsarticle}
\usepackage{amssymb}
\usepackage{amsmath}
\journal{Icarus}
\usepackage{lineno}
\hfuzz=\maxdimen
\vfuzz=\maxdimen
\usepackage{subcaption}
\usepackage{makecell} 

\usepackage{xcolor}
\usepackage[citebordercolor=white]{hyperref}
\hypersetup{colorlinks=true,linkcolor={blue},citecolor={blue},urlcolor={red}}

\begin{document}

\begin{frontmatter}

\title{A study on the contribution of the interplanetary medium in radio occultation experiments} 
\author[label1]{Keshav Aggarwal\corref{cor1}}
\author[label2]{R. K. Choudhary}
\author[label1]{Abhirup Datta}
\author[label2]{Soumyaneal Banerjee}
\author[label3]{R. Manikantan}
\author[label3]{Anshuman Sharma} 
\author[label4]{Takeshi Imamura}
\affiliation[label1]{organization={Department of Astronomy, Astrophysics and Space Engineering (DAASE), Indian Institute of Technology Indore},
            addressline={Khandwa Road, Simrol},
            city={Indore},
            postcode={453552},
            state={Madhya Pradesh},
            country={India}}
\affiliation[label2]{organization={Space Physics Laboratory (SPL), Indian Space Research Organization},
            addressline={Vikram Sarabhai Space Centre},
            city={Thiruvananthapuram},
            postcode={695022},
            state={Kerala},
            country={India}}

\affiliation[label3]{organization={ISRO Telemetry Tracking and Command Network (ISTRAC)},
            city={Bengaluru},
            postcode={560058},
            state={Karnataka},
            country={India}}
\cortext[cor1]{Corresponding author: keshavagg1098@gmail.com}

\affiliation[label4]{organization={Graduate School of Frontier Sciences, The University of Tokyo},
            addressline={Kiban-tou 4H7, 5-1-5 Kashiwanoha},
            city={Kashiwa},
            postcode={277-8561},
            state={Chiba},
            country={Japan}}
\begin{abstract}
Irregularities in electron density within the interplanetary medium (IPM) can cause fluctuations in the Doppler frequency of spacecraft radio signals. The amplitude of these fluctuations depends on factors such as the carrier frequency, propagation geometry, and link configuration. However, quantitative characterization of these effects across different frequencies in various occultation experiments is currently limited. We analyze five complementary datasets: two-way S-band observations from Chandrayaan-3 outside the lunar ionosphere, two-way S-band data from Chandrayaan-2 during lunar occultation, one-way S/X band measurements from the Venus Express Radio Science (VeRa)/Akatsuki Radio Science (Akatsuki) under IPM-only conditions, and one-way X-band Akatsuki data during solar occultation. The Chandrayaan-3 and Akatsuki IPM observations isolate IPM effects by excluding contributions from planetary atmospheres, the lunar ionosphere, and, except during solar occultation, the solar corona. Chandrayaan-3 data sample dynamically evolving Earth-Moon geometries and exhibit weak, mHz-level Doppler fluctuations, while Chandrayaan-2 observations provide near-lunar plasma benchmarks with higher amplitudes, during quiet time solar and geomagnetic conditions. Akatsuki and VeRa's IPM-only measurements capture long-path interplanetary effects, whereas Akatsuki solar occultation data reveal strong coronal signatures. Power spectral density analysis indicates Kolmogorov-like turbulence for lunar occultation and solar occultation cases, while IPM-only spectra show low-amplitude fluctuations. These results quantify the IPM contribution to Doppler noise, demonstrate the enhanced plasma sensitivity of two-way coherent links, and provide constraints relevant to turbulence modelling, precision spacecraft tracking, and the interpretation of radio occultation experiments.
\end{abstract}
\begin{keyword}
Radio occultation \sep Solar wind \sep Interplanetary medium
\end{keyword}
\end{frontmatter}
\section{Introduction}

The interplanetary medium (IPM) is a dilute, magnetized plasma region between Sun, planets and the interstellar medium, formed by the continuous expansion of the solar corona into the heliosphere, commonly known as the solar wind. Composed primarily of electrons and protons and permeated by magnetic fields of solar origin, the IPM governs the transport of mass, momentum, and energy throughout the heliosphere. Theoretical predictions of a steady solar outflow were first formulated by \citet{Parker1958} and were soon confirmed by direct spacecraft measurements \citep{Gringauz1960}. Since then, a substantial body of observational and theoretical work has been devoted to characterizing plasma properties close to the Sun, within planetary magnetospheres, and at large heliocentric distances. In contrast, comparatively fewer studies have focused on the IPM at spatial scales comparable to Earth-planet and Earth-Moon separations, despite the importance of this regime for heliospheric science and precision radio tracking of spacecraft.

Electromagnetic waves propagating through the IPM interact with free electrons along the propagation path, resulting in dispersive delays, phase scintillations, and Doppler frequency fluctuations in the received signals. These effects depend on both the total electron content integrated along the line of sight (LOS) and on stochastic fluctuations in electron density associated with plasma turbulence \citep{Woo1979, Coles1978, Armstrong1981, Aggarwal2025a, Aggarwal2025b, Aggarwal2026, Aggarwal2026b}. For a cold plasma, the refractive index scales inversely with the square of the radio frequency, causing lower-frequency signals to be more strongly affected by plasma-induced perturbations. Consequently, S-band radio signals exhibit significantly greater sensitivity to IPM effects than higher-frequency signals such as X-band \citep{Woo1979, Efimov2002}. This frequency dependence has been widely exploited in solar conjunction experiments to probe solar wind turbulence, coronal density structure, and plasma irregularities when radio signals pass close to the Sun \citep{Muhleman1977, Paetzold1996, Efimov2005, Efimov2010}.

Radio occultation (RO) and precision radio tracking experiments provide an effective remote-sensing approach for investigating plasma environments by monitoring perturbations in spacecraft radio signals \citep{Woo1979, Bird1982, Efimov2002, Armand2003, Imamura2005, Ando2015, Tripathi2022b, wexler2022, Aggarwal2026, Aggarwal2026b}. When the LOS intersects dense plasma regions, such as planetary atmospheres, ionospheres, or the solar corona, the resulting signal fluctuations are dominated by these localized structures. Conversely, when the LOS avoids planetary atmospheres, the lunar ionosphere, and regions near the Sun, the observed fluctuations primarily reflect the background IPM. Such observing geometries enable the study of interplanetary plasma irregularities over spatial scales that are difficult to probe using in situ measurements alone, particularly in regions where spacecraft coverage is sparse.

Previous radio occultation studies have focused on solar and planetary conjunction geometries, where coronal plasma and planetary atmosphere dominate the signal perturbations \citep{Woo1979, Coles1991, Fjeldbo1971}. By comparison, relatively few studies have examined radio signals propagating along paths that are well separated from both the solar corona and planetary plasma environments \citep{Calves2014, Tripathi2022b}. These non-conjunction observations are essential for isolating the intrinsic contribution of the IPM and for improving models of plasma-induced noise in high-precision applications, including deep-space navigation, gravitational experiments, and planetary radio science.

Recent planetary missions provide new opportunities to address this observational gap. The Chandrayaan-2 (CH2) mission demonstrated the use of two-way coherent S-band radio occultation measurements to study plasma-induced Doppler fluctuations in the near-lunar environment \citep{Tripathi2025}. Building on this capability, the Chandrayaan-3 propulsion module (CH3PM) employed a similar two-way coherent S-band system but operated under a markedly different observing geometry. In contrast to the relatively stable lunar orbit of CH2, CH3PM followed a highly elliptical trajectory influenced by both Earth’s and Moon’s gravity, which gave us the opportunity to ensure that the signal path remained free of the lunar ionosphere and allowed measurements dominated by interplanetary plasma effects.

During its second lunar flyby on 11 November 2025, CH3PM reached a closest approach of approximately 4,537 km from the lunar surface. The gravitational influences of the Earth and Moon significantly altered the spacecraft's trajectory, resulting in Earth-Moon line-of-sight distances ranging from approximately 30,000 km to 90,000 km from the Moon. This observation interval also coincided with enhanced solar activity, including multiple X-class solar flares and coronal mass ejections, leading to moderate to strong disturbances in the IPM. The combination of dynamically evolving propagation geometry and elevated solar activity provides a rare opportunity to examine IPM-induced radio fluctuations under conditions that differ substantially from typical conjunction experiments.

Complementary observations are provided by the Venus Orbiter Akatsuki (Akatsuki), which offers X-band open-loop radio-tracking data during both solar occultation experiments and intervals when Venus and Earth are on the same side of the Sun \citep{Imamura2011, Imamura2017, Ando2015, Jain2023}. Although X-band signals are intrinsically less sensitive to plasma-induced effects due to their higher frequency, in this study, they traverse significantly longer interplanetary paths. In this study, Akatsuki observations are selected to avoid Venus’s atmosphere and the solar corona, thereby isolating the IPM contribution. We also use observations from the Venus Express mission to compare the effect of the IPM on S-band signals. Solar occultation measurements are additionally used to assess plasma-induced effects at smaller solar offsets and under varying levels of solar activity.

A combined analysis of S-band and X-band observations acquired under comparable interplanetary medium conditions for Venus-Earth and Moon-Earth systems enables a direct assessment of how radio frequency, propagation geometry, and system configuration influence sensitivity to IPM-induced fluctuations. In this study, the interplanetary medium conditions along the line of sight (LOS) between the spacecraft and Earth include the background electron density, the level of plasma turbulence, solar wind conditions, heliocentric distance and solar offset of the LOS, and the effective propagation path length. For the observations considered here, these parameters correspond to quiet solar and geomagnetic conditions, as indicated by the Dst values listed in Table\ref{tab:doppler_summary}, ensuring that the measured Doppler fluctuations primarily reflect intrinsic interplanetary plasma effects rather than transient space weather disturbances.Two-way coherent systems, such as those employed by CH2 and CH3PM, enhance plasma signatures because the radio signal traverses the same path twice, causing plasma-induced phase perturbations to add constructively while many instrumental noise sources cancel out \citep{Woo1979}. In contrast, one-way open-loop systems, such as Venus Express Radio Science Experiment (VeRa) S-band and Akatsuki X-band trackings, do not benefit from this cancellation but provide valuable complementary measurements at higher frequencies and over extended baselines.

The primary objectives of this study are to quantify Doppler frequency fluctuations induced by the IPM using high-stability S-band and X-band spacecraft radio signals, to compare these fluctuations under similar geomagnetic and solar conditions, and to assess the influence of orbital geometry and LOS distance on the observed plasma-induced variations. All observations analyzed in this study correspond to relatively quiet solar and geomagnetic conditions, as indicated by the daily averaged Dst values listed in Table \ref{tab:doppler_summary}. This minimizes the influence of transient space weather events such as geomagnetic storms or strong solar wind disturbances. Therefore, the observed variations in Doppler fluctuations and spectral indices are primarily attributed to differences in the plasma environment along the line of sight and propagation geometry, rather than external driving by solar activity. By combining Chandrayaan-3 propulsion module two-way coherent S-band observations, CH2 S-band radio occultation data, VeRa S-band observations and Akatsuki X-band open-loop tracking data, this work provides a multi-frequency characterization of IPM effects on spacecraft radio propagation and establishes observational constraints relevant for future deep-space radio science and navigation experiments.

\section{Method}
This study examines the impact of the interplanetary medium (IPM) on spacecraft radio signals, with a focus on Doppler frequency fluctuations caused by plasma. The IPM is a magnetized plasma composed of electrons and protons carried outward from the Sun; as a result, radio waves traversing it experience dispersive delays, phase fluctuations, and Doppler shifts. Understanding these effects is critical for high-precision spacecraft tracking, radio science experiments, and studies of plasma turbulence far from the Sun.

We analyze five distinct datasets, each chosen to probe different plasma contributions along the line-of-sight (LOS) path:

\begin{itemize}
    \item \textbf{CH3PM S-band in IPM:} Observations from 10 November 2025 (DOY 314) capture the Chandrayaan-3 propulsion module well outside the lunar atmosphere. The Earth-Moon LOS distance during this interval is approximately 35,000 km, with a highly elliptical orbit that produces large variations in geometry. These data isolate the interplanetary contribution to S-band Doppler fluctuations, free from lunar ionosphere effects.

    \item \textbf{CH2 S-band in lunar ionosphere:} Observations from 08 November 2022 (DOY 312) occur when the Moon and CH2 orbiter are inside the terrestrial magnetotail. These data capture both lunar ionosphere and near-lunar plasma contributions. This dataset provides a reference for typical S-band Doppler residuals in lunar radio occultation studies and allows direct comparison against CH3PM to show how IPM effects are smaller than near-lunar plasma effects.

    \item \textbf{Venus Express S-band in IPM:} Observations from 15 Dec 2008 capture the signals from the probe when the LOS path was $\sim 0.91$ AU. These data isolate the interplanetary contribution at S-band frequency ($\sim$2.3 GHz) and show Doppler fluctuations of the order $\pm$2.55 Hz over long propagation distances.

    \item \textbf{Akatsuki X-band in IPM:} Observations from 08 June 2020 capture the signals from the probe when the LOS path between Earth and Venus are at their closest ($\sim 0.29$ AU). These data isolate the interplanetary contribution at X-band frequency ($\sim$8.41 GHz) and show Doppler fluctuations of the order $\pm$0.05 Hz over long propagation distances.

    \item \textbf{Akatsuki X-band in solar occultation:} Observations from 23 October 2022, when the LOS passes within 3.9 $R_{\odot}$ of the solar center, capture the contribution of the solar corona and solar wind. These data show the extreme enhancement of Doppler fluctuations at small solar offsets, up to $\pm$2 Hz, demonstrating the dominant effect of the coronal plasma on spacecraft radio signals.

\end{itemize}

\subsection{Chandrayaan-3 S-band Observations (IPM)}

The Chandrayaan-3 propulsion module (CH3PM) uses a two-way coherent S-band radio link with an uplink frequency of 2028.78 MHz and a fixed turnaround ratio (TAR) of 240/221, referenced to an ensemble of Cesium and MASER clock sources at the Indian Deep Space Network (IDSN). The TAR is a predetermined numerical ratio of the frequency transmitted by a spacecraft's transponder to the frequency it receives from a ground station in a two-way radio science experiment. As a result, the downlink inherits the phase stability of the ground-based reference, and the onboard oscillator noise is strongly suppressed.

Open-loop recordings at IDSN provide high-cadence measurements of the received downlink frequency. These data include contributions from spacecraft and Earth motion, relativistic effects, and refractive delays introduced by all media through which the signal passes. The primary observable in this study is the received coherent downlink frequency, $f_{\text{obs}}(t)$. Deterministic Doppler contributions associated with spacecraft motion, Earth rotation, and relativistic light-time effects are removed by computing a predicted frequency, $f_{\mathrm{th}}(t)$, using a relativistic light-time model analogous to that employed by \citet{Tripathi2025}, which can be estimated using the SPICE toolkit \citep{Acton1996, Annex2020}. The Doppler frequency residuals are defined as

\begin{equation}
\Delta f(t) = f_{\text{obs}}(t) - f_{\mathrm{th}}(t)
\end{equation}
representing the cumulative refractive effects of plasma irregularities along the two-way propagation path. The analysis presented here is based on observations acquired on 10 November 2025 (DOY 314), when CH3PM was well outside the lunar atmosphere. This dataset was chosen because it offers the longest continuous tracking interval free from the influence of spacecraft attitude parameter evolution. Moreover, at such large distances from the Moon, the line of sight sampled a relatively uniform plasma environment, with only minor variations in the propagation geometry. As a result, the spectral characteristics, including the spectral slopes, are expected to remain nearly invariant over the observation period.

During the period of the observation, the Earth-Moon LOS distance was approximately 35,000 km from the lunar center. The spacecraft followed a highly elliptical orbit influenced by both Earth and Moon gravity, which caused the LOS geometry and propagation distance to vary significantly over the observation interval.
These data allowed us to isolate the interplanetary contribution to S-band Doppler residuals, as the lunar ionosphere is absent. Solar activity during this period, including multiple X-class flares and coronal mass ejections (CMEs), produced disturbances in the IPM, allowing study of plasma-induced fluctuations under dynamically changing conditions. 

\begin{figure*}[ht!]
\centering
\includegraphics[width=\linewidth]{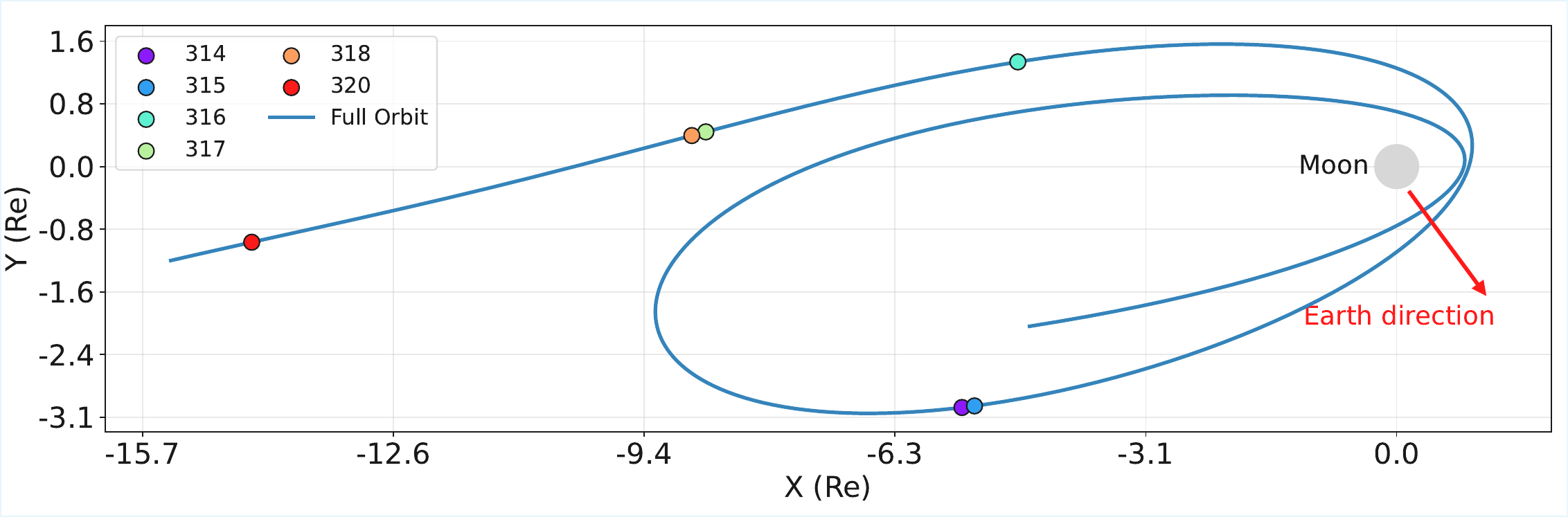}
\caption{Positions of the CH3PM propulsion module during DOY 313–320 2025. Blue curve: spacecraft trajectory; colored points: positions; red arrow: mean Earth direction. This geometry provides a range of LOS distances between 30,000–90,000 km, with DOY 314 used for detailed Doppler analysis.}
\label{fig:CH3POS}
\end{figure*}

Figure \ref{fig:CH3POS} shows the positions of the CH3PM propulsion module during 10th to 16th November 2025 (DOY 314–320). The blue curve shows the spacecraft trajectory, and the colored points show the daily positions of the module. The red arrow marks the mean Earth direction. This geometry provides a range of LOS distances between 30,000–90,000 km, with DOY 314 ($\sim 30,000 km$) used for detailed Doppler analysis.

\subsection{Chandrayaan-2 S-band Observations (Lunar Ionosphere)}

The Chandrayaan-2 (CH2) orbiter provides a two-way coherent S-band link with an uplink frequency of 2041.598 MHz and a fixed turnaround ratio (TAR) of 240/221. We analyze data from 08 November 2022 (DOY 312), when the Moon and CH2 orbiter were inside the magnetotail region. Unlike CH3PM, the CH2 orbital configuration is relatively stable with a circular orbit of $100 \times 100$ km, and the LOS distance between Earth and the spacecraft varies minimally. 
These observations pass through the lunar ionosphere, capturing Doppler fluctuations caused by near-lunar plasma. The two-way coherent configuration enhances plasma-induced phase fluctuations by accumulating them along both uplink and downlink, while canceling many ground- and spacecraft-related noise sources. The CH2 dataset serves as a reference for typical S-band Doppler residuals observed in lunar radio occultation studies and provides a benchmark for comparing with the CH3PM data, which contains only IPM contributions.

\begin{figure*}
\centering
\includegraphics[width=0.45\linewidth]{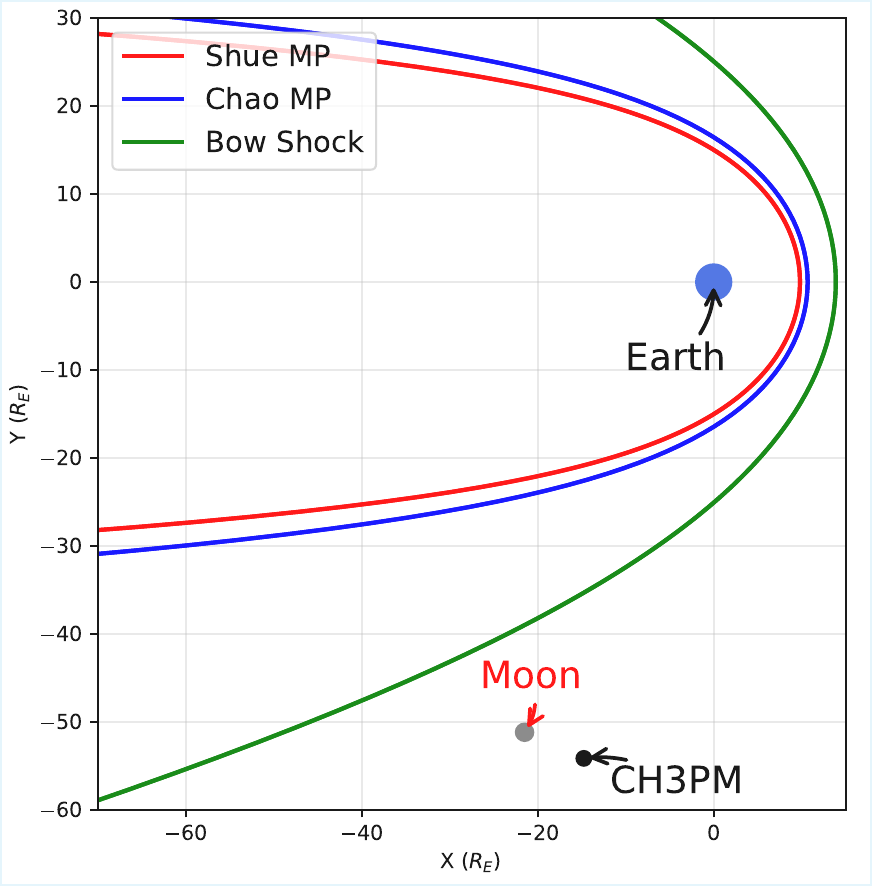}
\includegraphics[width=0.5\linewidth]{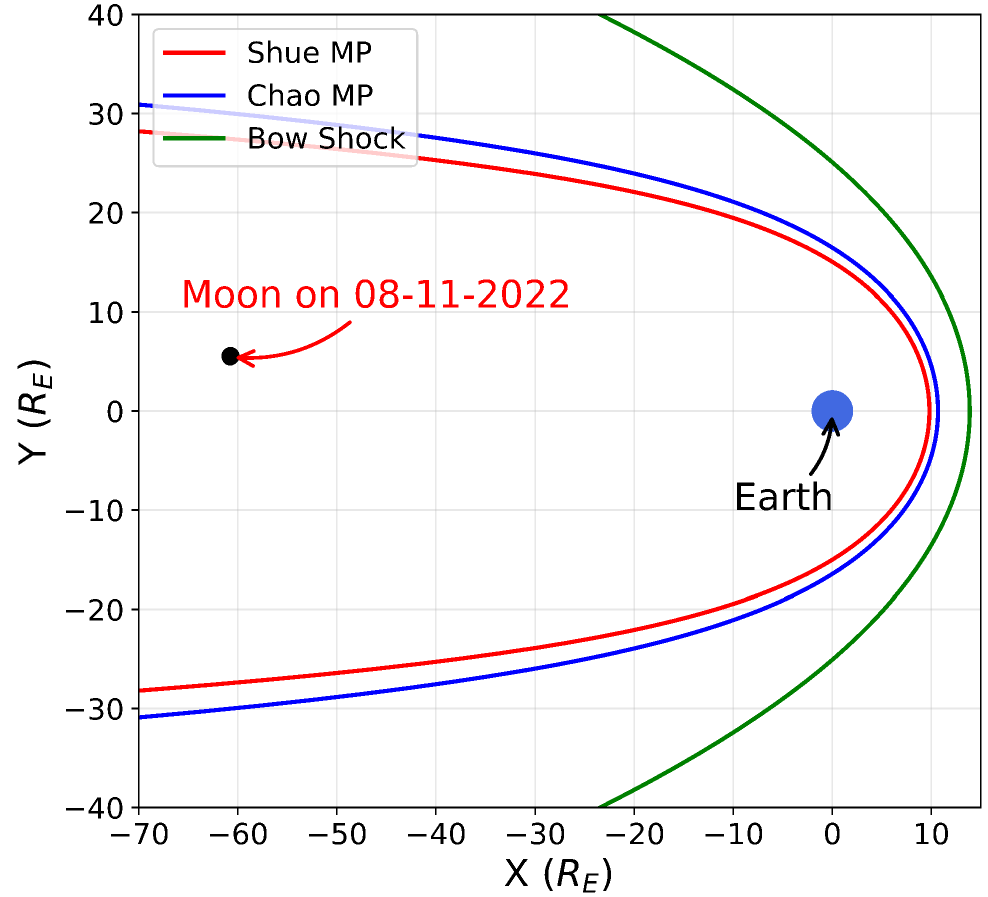}
\caption{\textbf{Left Panel:} Position of the Moon (point in grey) and CH3PM (point in black) on 10 November 2025 (DOY 314) 2025. \textbf{Right panel:} Position of the Moon/CH2 orbiter on 08 November 2022 (DOY 312, black) against the position of Earth (blue point). This configuration allows Doppler residuals to include contributions from lunar ionosphere and near-lunar plasma. Theoretical magnetopause boundaries are shown in red and blue, and the bow shock boundary in green for both panels \citep{Shue1997, Chao2002}.}
\label{fig:ch2_pos}
\end{figure*}

Figure \ref{fig:ch2_pos} shows the position of the Moon/CH2 orbiter on 08 November 2022 (DOY 312, black) against the position of Earth (blue point). Theoretical magnetopause boundaries are shown in red and blue, and the bow shock boundary in green, following \cite{Shue1997, Chao2002}. This configuration allows Doppler residuals to include contributions from lunar ionosphere and near-lunar plasma.

\subsection{Venus Express S-band Observations (IPM)}
\begin{figure}[ht!]
\centering
\includegraphics[width=\linewidth]{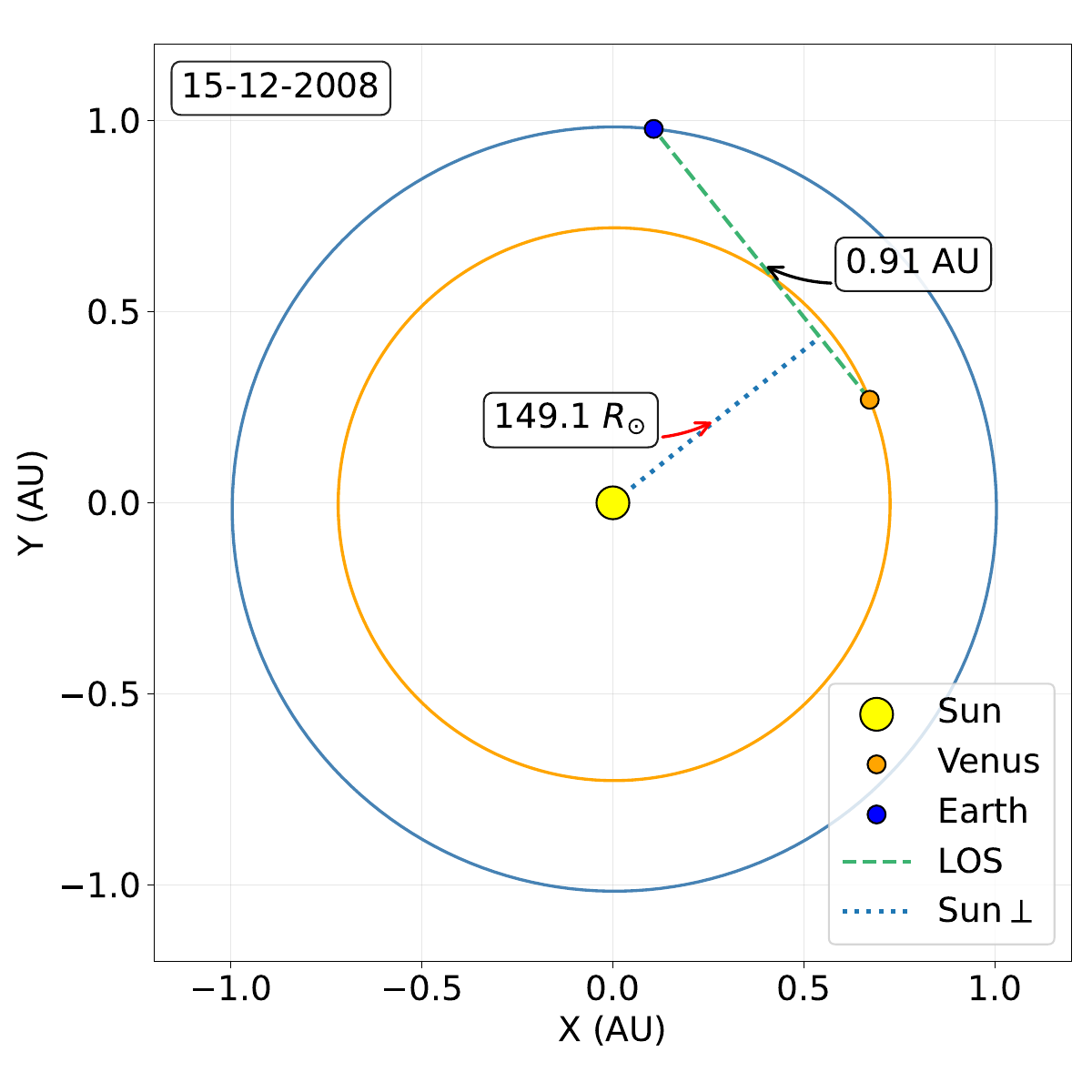}
\caption{Observing geometry for VeRa S-band measurements on 15 Dec 2008. Sun: yellow, Earth: blue, Venus: orange. Green dashed line: signal path; blue dotted line: closest approach to Sun (red arrow).}
\label{fig:VEX_positions}
\end{figure}

We utilized observations made by the Venus Express probe on 15th Dec 2008 which were received at the New Norcia ground station, Australia, in the S band, which enabled us to investigate the extent of the IPM's contribution to these signals typically used for Solar occultation and Venus occultation studies. Additional information about data acquisition can be found in other sources \citep{Husler2006}. We process the closed-loop S-band ($\sim$2.3 GHz) downlink data received from the VEX probe using the method described above, to estimate the residuals which we use for this study. Figure \ref{fig:VEX_positions} shows the observing geometry for VeRa S-band measurements.

\subsection{Akatsuki X-band Observations (IPM and Solar Occultation)}
We also utilized observations made by the Japanese Akatsuki probe, received at the Usuda Deep Space Center (UDSC) in Japan, in the X band, which enabled us to investigate the extent of the IPM's contribution to these higher-frequency signals, typically used for Solar occultation and Venus occultation studies. Additional information about data acquisition can be found in other sources \citep{Imamura2011, Tripathi2023}. We process the data received from the Akatsuki probe using the method described above, to estimate the residuals which we use for this study. The Akatsuki spacecraft around Venus provides open-loop X-band ($\sim$8.41 GHz) downlink data in two different conditions:

\begin{itemize}
    \item \textbf{IPM-only:} Observed on 08 June 2020, when the Earth-Venus LOS is maximally separated from the Sun, ensuring the observed Doppler fluctuations are dominated by interplanetary plasma along the long propagation path.
    \item \textbf{Solar occultation:} Observed on 23 October 2022, when the LOS passes within 3.9 $R_{\odot}$ of the solar center. Doppler fluctuations are greatly enhanced, showing the dominant contribution of the solar corona and solar wind at small heliocentric offsets.
\end{itemize}

\begin{figure*}[!h]
\centering
\includegraphics[width=\linewidth]{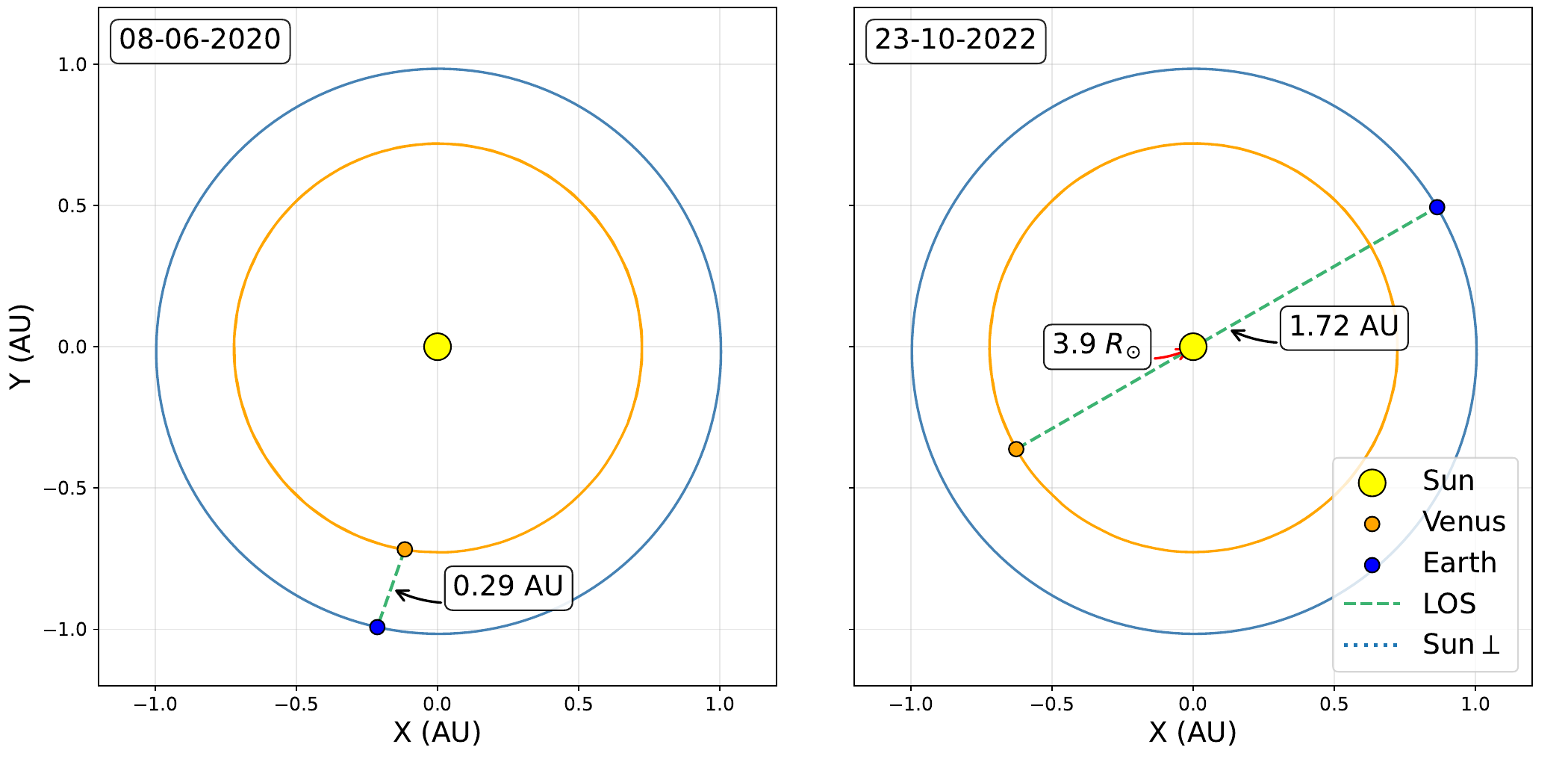}
\caption{Observing geometry for Akatsuki X-band measurements. Left: Maximum LOS separation from the Sun on 08 June 2020 (IPM-only). Right: Solar occultation on 23 October 2022 (3.9 $R_{\odot}$). Sun: yellow, Earth: blue, Venus: orange. Green dashed line: signal path; blue dotted line: closest approach to Sun (red arrow).}
\label{fig:Akatsuki_positions}
\end{figure*}

Figure \ref{fig:Akatsuki_positions} shows the observing geometry for Akatsuki X-band measurements. Left panel shows the maximum LOS separation from the Sun on 08 June 2020 (IPM-only). Right panel shows the geometry of the Solar occultation experiment on 23 October 2022 (3.9 $R_{\odot}$). The Sun is marked in yellow, the Earth is marked in blue, and Venus is shown in orange. The green dashed line shows the signal path/ LOS path and the blue dotted line shows the distance of closest approach to Sun (red arrow).

These five datasets allow comparisons across different frequencies, orbital geometries, plasma environments, and solar activity levels. CH3PM, VEX and Akatsuki IPM datasets isolate interplanetary contributions, whereas CH2 and Akatsuki solar occultation datasets capture the influence of near-lunar and coronal plasma. This design enables us to examine how radio frequency, LOS distance, and plasma environment affect Doppler fluctuations.

\section{Results and Discussion}

\subsection{S-band Doppler Residuals: CH3PM and CH2}

\begin{figure*}[ht!]
\centering
\includegraphics[width=\linewidth]{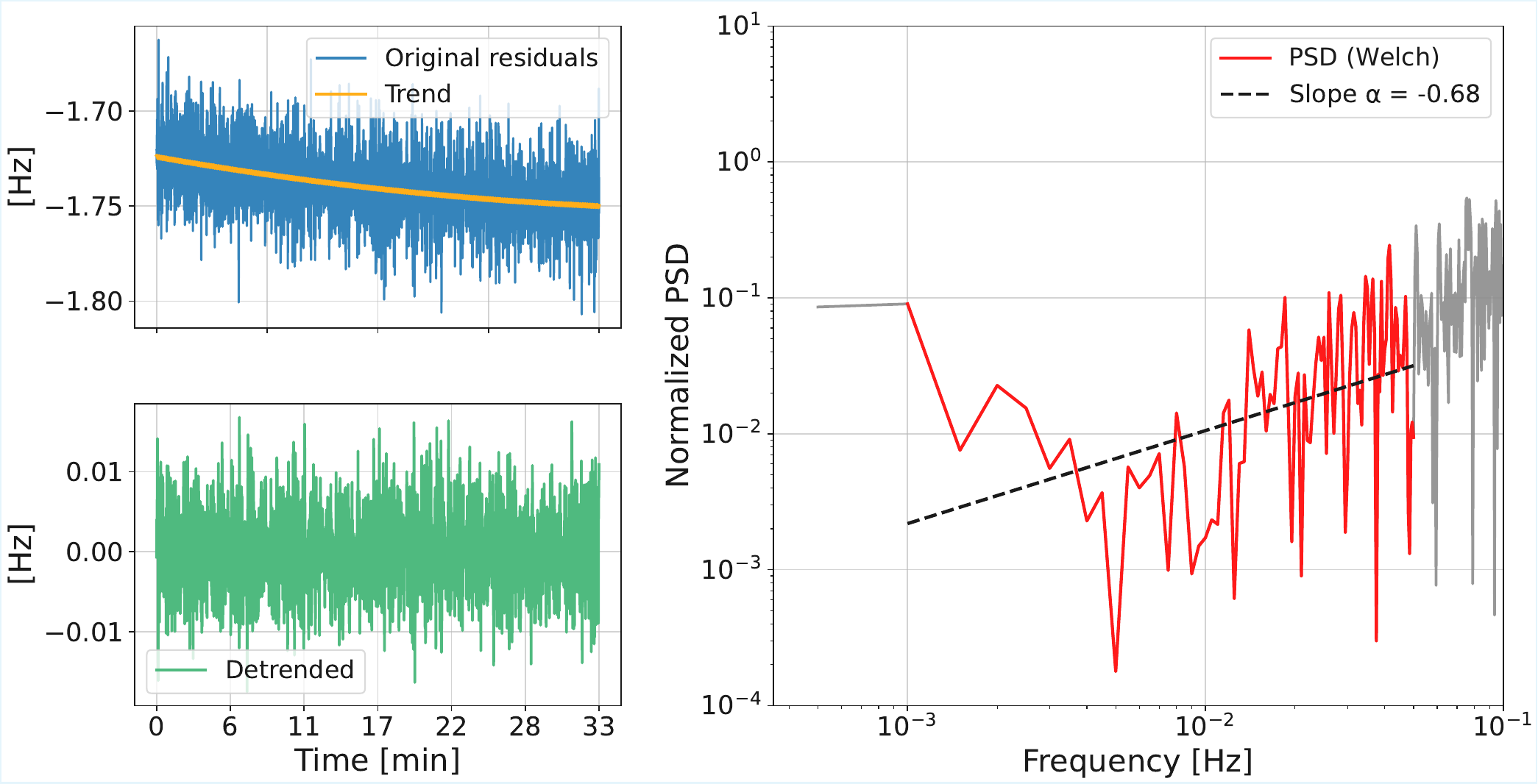}
\caption{CH3PM S-band observations for DOY 314, 10 November 2025. Top left: raw Doppler residuals (blue) with trend fit (orange). Bottom left: detrended residuals showing slow fluctuations attributed to the IPM. Right: PSD of the detrended residuals using the Welch method (red), slope fit in black dashed.}
\label{fig:signal_processing}
\end{figure*}

The CH3PM S-band Doppler residuals for DOY 314 (10 November 2025) are shown in Figure \ref{fig:signal_processing}. The residuals are shown in blue in top left panel, with the trend fit in orange to highlight the general behavior of the signal. In the bottom left panel, the detrended residuals are presented in green. On the right, the power spectral density (PSD) is plotted in red, with the slope fit in the range defined by $[1/N, f_N/10]$ indicated by a black dashed line. Here, N is the sample length, and $f_N$ is the Nyquist frequency of the signal. The spacecraft was outside the lunar ionosphere, so the observed fluctuations primarily reflect the Moon-Earth IPM contribution. The raw residuals show slow variations over the observation interval, which are detrended using a second-order polynomial \citep{Aggarwal2026, Wexler2019, Gramigna2023}. After detrending, most of the residuals lie within $\pm$0.01 Hz. Persistent mHz-level fluctuations indicate measurable but weak IPM effects. This dataset illustrates the effect of the IPM on the S-band signal, excluding contamination from the lunar ionosphere.

\begin{figure*}[ht!]
\centering
\includegraphics[width=\linewidth]{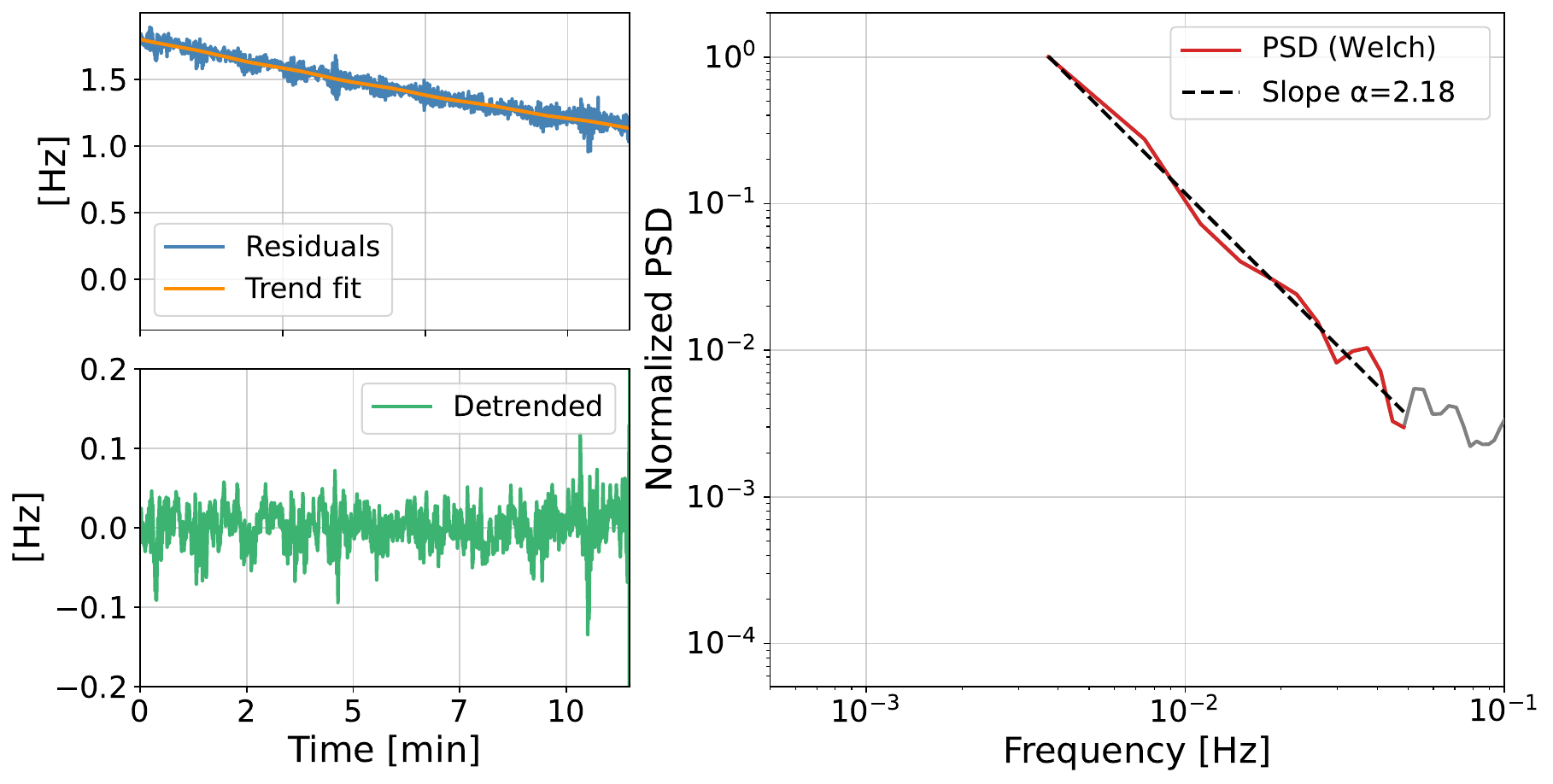}
\caption{CH2 S-band Doppler residuals (DOY 312, 08 November 2022). Top left: raw Doppler residuals (blue) with composite fit (orange). Bottom left: detrended residuals (green), showing contributions from lunar ionosphere and near-lunar plasma. Right: PSD of detrended residuals using the Welch method (red), slope fit in red dashed.}
\label{fig:ch2_resi}
\end{figure*}

For comparison, CH2 S-band Doppler residuals from DOY 312 (08 November 2022) are shown in Figure \ref{fig:ch2_resi}. The residuals are shown in blue in top left panel, with the trend fit in orange to highlight the general behavior of the signal. In the bottom left panel, the detrended residuals are presented in green. On the right, the power spectral density (PSD) is plotted in red, with the slope fit in the same range as above, indicated by a red dashed line. These observations include contributions from the lunar ionosphere and near-lunar plasma. Residuals range $\pm$0.075 Hz, significantly larger than CH3PM, reflecting the stronger plasma influence near the Moon. The two-way coherent configuration amplifies plasma-induced phase fluctuations, providing a benchmark for typical S-band Doppler perturbations in lunar occultation studies.

\subsection{S-band Doppler Residuals: VeRa}

\begin{figure*}[ht!]
\centering
\includegraphics[width=\linewidth]{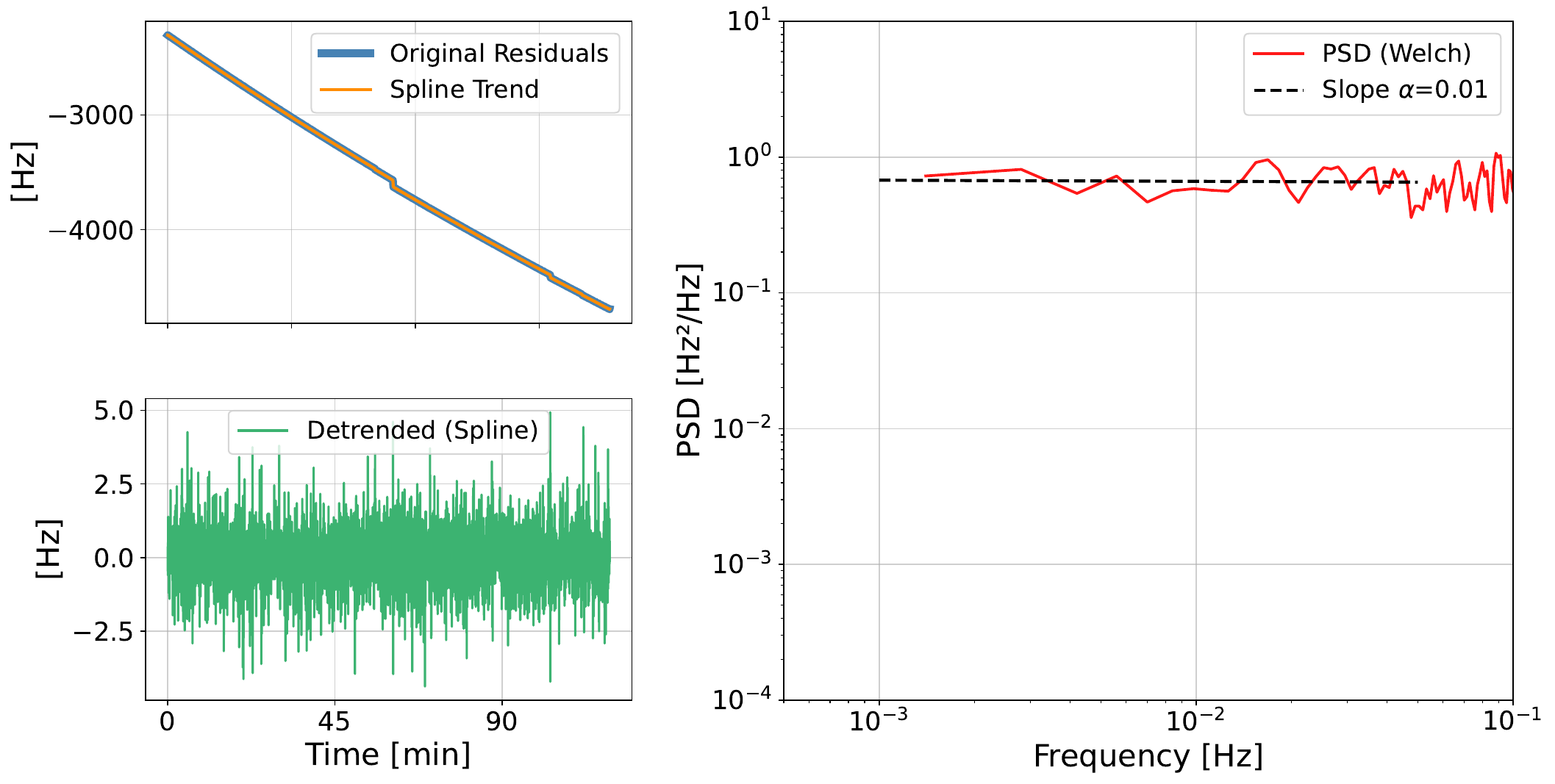}
\caption{VeRa S-band IPM-only residuals (15 Dec 2008). Top left: raw residuals (blue) with trend (orange). Bottom left: detrended residuals showing slow fluctuations attributed to IPM. Right: PSD using the Welch method (red) with slope fit (black dashed).}
\label{fig:VEX_ipm}
\end{figure*}

VeRa S-band Doppler residuals in the IPM-only configuration (15 Dec 2008 - DOY 350) are shown in Figure \ref{fig:VEX_ipm}. The residuals are shown in blue in top left panel, with the trend fit in orange to highlight the general behavior of the signal. In the bottom left panel, the detrended residuals are presented in green. On the right, the power spectral density (PSD) is plotted in red, with the slope fit in the same range, indicated by a red dashed line. The majority of the residuals lie within $\pm$2.5 Hz, demonstrating the measurable effect of the IPM on S-band signals.

\subsection{X-band Doppler Residuals: Akatsuki}

\begin{figure*}[ht!]
\centering
\includegraphics[width=\linewidth]{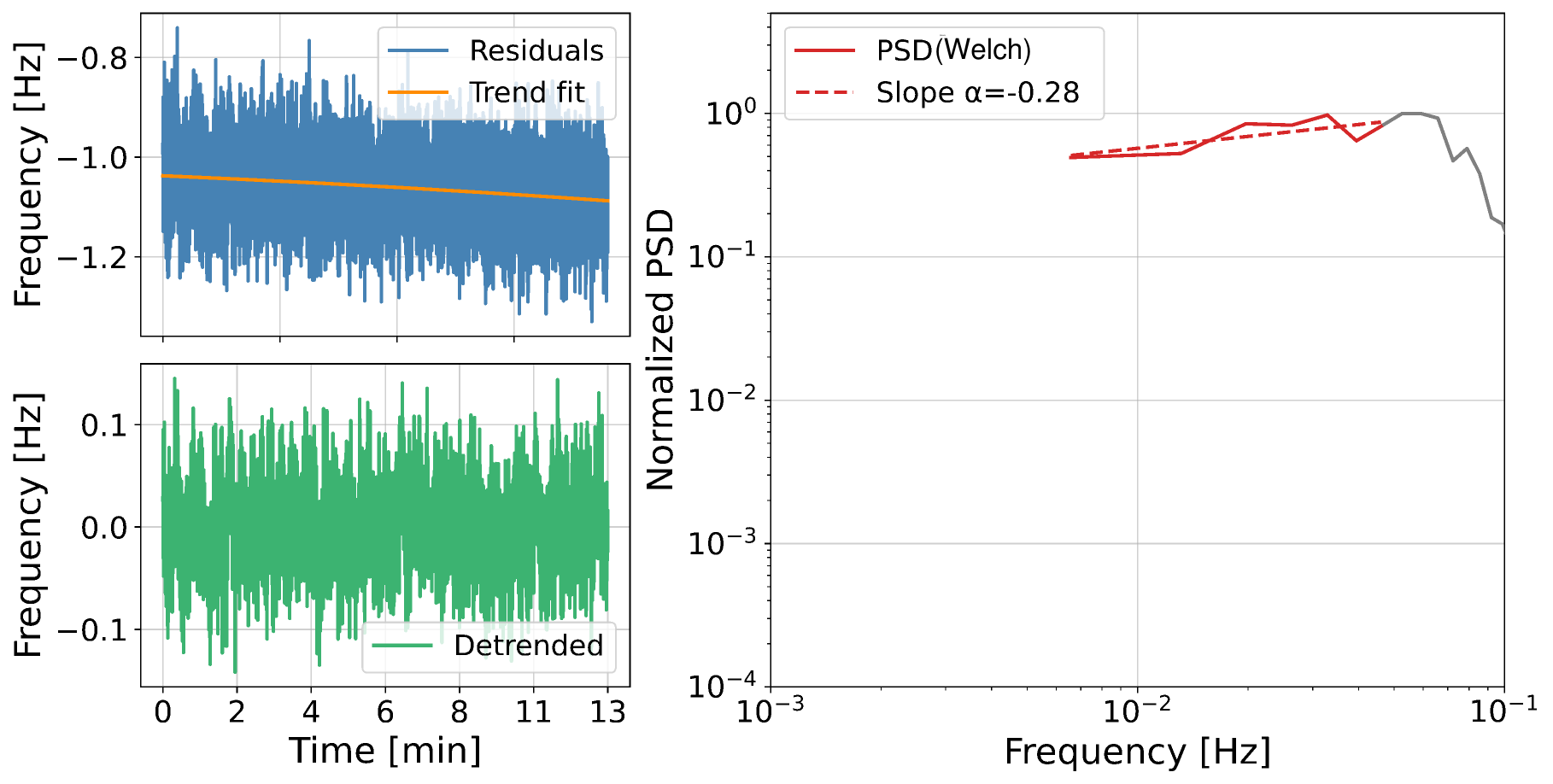}
\caption{Akatsuki X-band IPM-only residuals (08 June 2020). Top left: raw residuals (blue) with trend (orange). Bottom left: detrended residuals showing slow fluctuations attributed to IPM. Right: PSD using the Welch method (red) with slope fit (black dashed).}
\label{fig:Akatsuki_ipm}
\end{figure*}

\begin{figure*}[ht!]
\centering
\includegraphics[width=\linewidth]{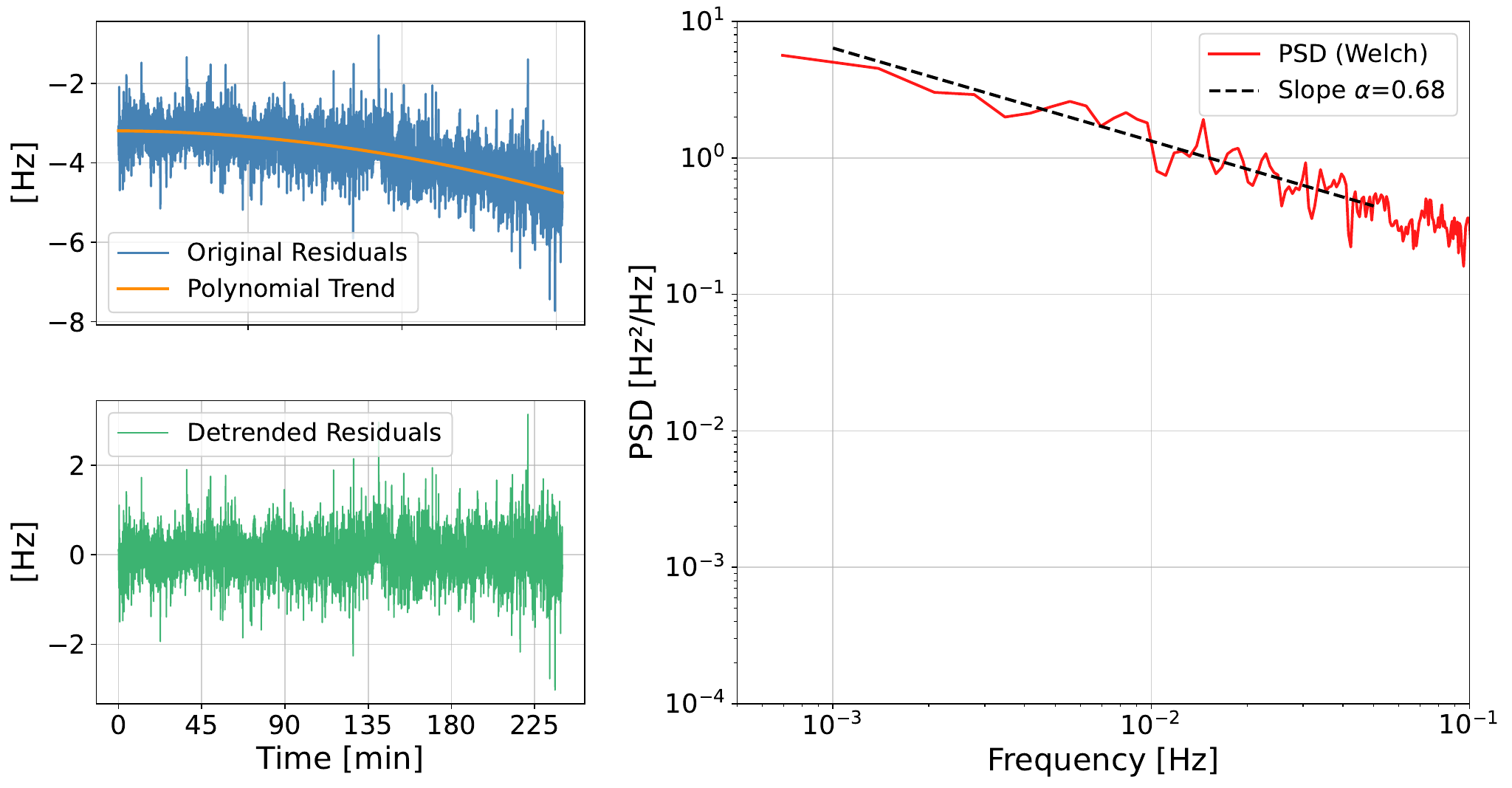}
\caption{Akatsuki X-band residuals during solar occultation (23 October 2022, 3.9 $R_{\odot}$). Top left: raw residuals (blue) with trend (orange). Bottom left: detrended residuals showing large fluctuations dominated by coronal plasma. Right: PSD using the Welch method (red) with slope fit (black dashed).}
\label{fig:Akatsuki_sol_resi}
\end{figure*}

Akatsuki X-band Doppler residuals in the IPM-only configuration (08 June 2020) are shown in Figure \ref{fig:Akatsuki_ipm}. The residuals are shown in blue in top left panel, with the trend fit in orange to highlight the general behavior of the signal. In the bottom left panel, the detrended residuals are presented in green. On the right, the power spectral density (PSD) is plotted in red, with the slope fit in the same range, indicated by a red dashed line. The majority of the residuals lie within $\pm$0.1 Hz, demonstrating the small but measurable effect of the IPM on high-frequency signals. In the solar occultation configuration (23 October 2022, 3.9 $R_{\odot}$), residuals increase dramatically, up to $\pm$2 Hz (Figure \ref{fig:Akatsuki_sol_resi}), illustrating the dominant effect of the solar corona and wind at small heliocentric offsets and reaffirm the approximation of negligible effects due to the IPM in such experiments.

\subsection{Comparison of RO Doppler Residuals with White Noise Models}

To better understand the spectral characteristics of the residual Doppler fluctuations observed in the CH3PM, CH2, VeRa and Akatsuki datasets, we generated a reference model using pure white Gaussian noise. In the code, we create a normalized white noise time series, scale it appropriately, and compute its Power Spectral Density (PSD) using Welch's method. This provided a baseline for comparison with the observed Doppler residuals. Unlike colored noise models, white noise has a flat PSD, serving as a reference to identify departures from purely random fluctuations in the measured data.

\begin{figure}[h!]
\centering
\includegraphics[width=\linewidth]{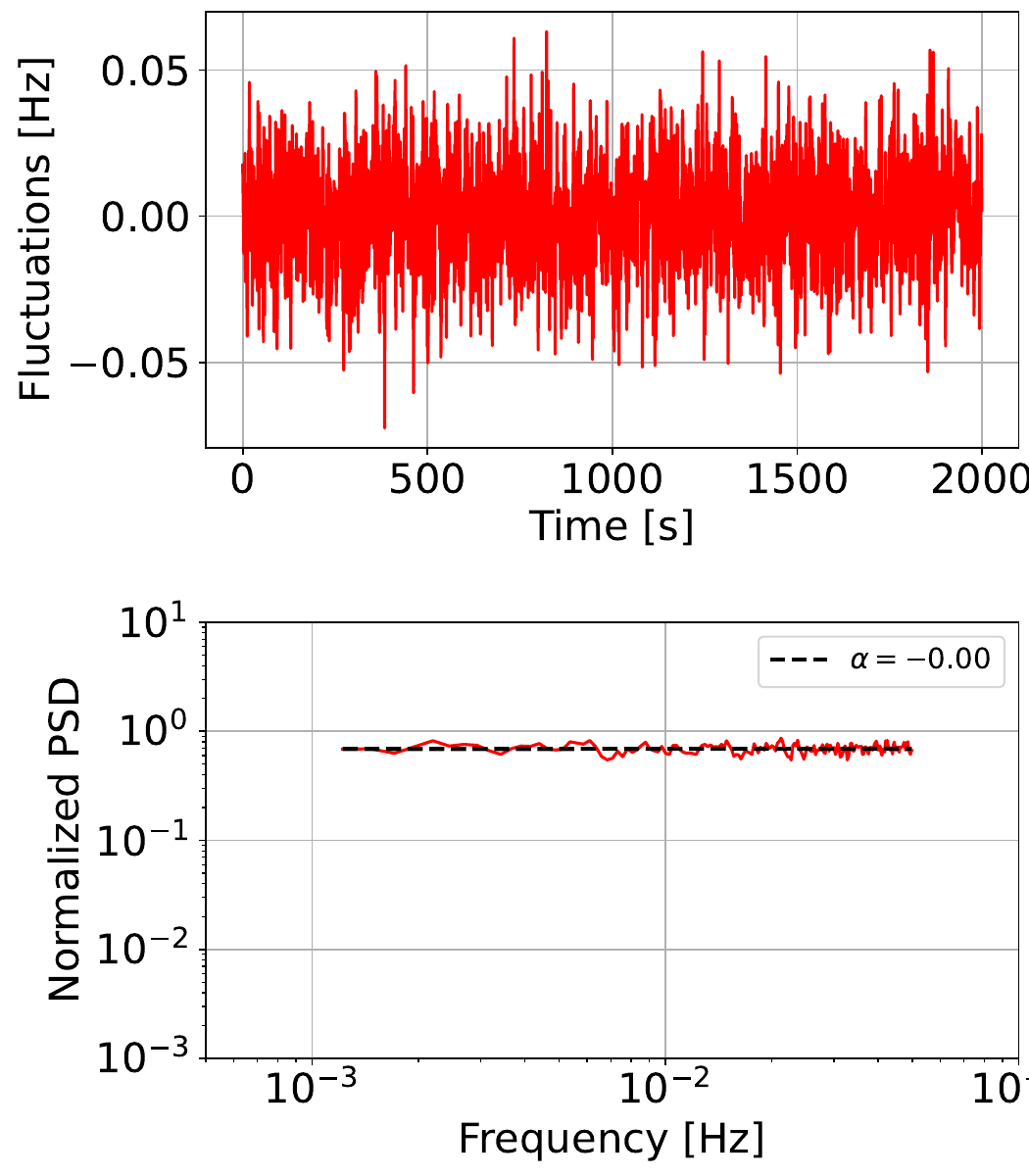}
\caption{Top Panel: Pure white noise generated using Python, with the amplitude of a similar order as seen in our signals in the IPM is shown in red. Bottom panel: Power Spectral Density of pure white noise generated for comparison with RO Doppler frequency residuals. The PSD is normalized and plotted over a selected frequency range to highlight its flat characteristic.}
\label{fig:psd_noise}
\end{figure}

The top panel in Figure \ref{fig:psd_noise} shows pure white noise generated using Python, with the amplitude of a similar order as seen in our signals in the IPM, is shown in red. The bottom panel shows the PSD of this noise for comparison with RO Doppler frequency residuals. The PSD of the generated white noise is normalized and plotted over the selected frequency range to highlight its flat characteristic. Comparing this with the PSDs of the RO Doppler residuals reveals that the observed fluctuations are not flat in frequency but exhibit frequency-dependent behavior. The deviation from the flat PSD of white noise indicates the presence of long-range correlations or structured variability in the plasma along the signal path.

The slope and structure of the measured PSDs provide quantitative insights into large-scale plasma variations in the interplanetary medium, lunar ionosphere, or solar corona. By contrasting with a white noise reference, we can separate purely random fluctuations from coherent or frequency-dependent effects, informing models of signal propagation, noise mitigation, and space weather analysis.

\subsection{Power Spectral Density Analysis and Interpretation}

Power spectral densities (PSDs) are computed for all five datasets to characterize the frequency-dependent behavior of plasma-induced Doppler fluctuations. For consistency, each PSD is normalized by the maximum value of power within that dataset, allowing direct comparison of relative fluctuation amplitudes across different spacecraft, frequencies, and observing conditions. The PSD analysis provides insight into both the amplitude and spectral distribution of plasma irregularities along the line-of-sight.

The measured PSD slopes provide quantitative insight into the nature of plasma turbulence along the line-of-sight. Following the convention $\alpha = p - 3$, with $p = 11/3$ for Kolmogorov turbulence, the expected slope is $\alpha = 2/3$ \citep{Armand2003, Efimov2003}. Positive slopes indicate dominance of large-scale fluctuations, while negative slopes correspond to no discernible information regarding the medium being traversed (as in the case of pure white noise shown in Figure \ref{fig:psd_noise}). In our datasets, Akatsuki solar occultation observations exhibit $\alpha \sim 0.68$, consistent with strong, energy-containing turbulence in the solar corona, while the observations by CH2 observation has $\alpha \sim 2.18$, which implies steep, non-Kolmogorov, dissipative turbulence.

CH3PM S-band, VeRa S-band, and Akatsuki IPM-only datasets have smaller or negative $\alpha$ (e.g., $-0.68$ to $0.01$), reflecting weak interplanetary turbulence far from the Sun and Moon. These findings align with previous radio and spacecraft measurements: near-Sun solar wind turbulence shows Kolmogorov-like slopes \citep{Efimov2010, Coles1989, Parker1958}, Venusian and Martian ionospheres exhibit steeper high-frequency slopes due to enhanced dissipation \citep{Kliore1979, Withers2009}, and Earth's magnetotail and magnetosheath regions show variable slopes depending on compressible MHD turbulence and solar wind driving \citep{Hadid2018, Tian2012}. Observations from Parker Solar Probe extend these measurements, revealing radial variations in slopes and the transition from fluid to kinetic-scale turbulence \citep{Lotz2023, Roberts2023, Badman2023}. Together, these comparisons demonstrate that PSD slopes not only quantify fluctuation amplitudes across scales but also serve as a diagnostic of turbulence strength, energy cascade, and plasma environment along the propagation path, from quiescent interplanetary space to strongly turbulent solar and planetary environments \citep{Rakhmanova2024, Jain2024}.

The CH2 S-band and Akatsuki X-band solar occultation datasets exhibit clear power-law behavior over a broad frequency range, with slopes consistent with Kolmogorov turbulence. Kolmogorov turbulence is characterized by a -5/3 power-law scaling in the inertial range of turbulent cascades, reflecting energy transfer from larger to smaller spatial scales in magnetized plasmas. In the context of Doppler measurements, the PSD slope quantifies how fluctuations vary with temporal frequency, providing an indirect probe of the underlying spatial scales of electron density irregularities. For CH2, the observed slope indicates that turbulence in the near-lunar plasma and magnetotail environment follows the expected cascade of turbulent energy, with significant contributions at both low and intermediate frequencies. Similarly, for Akatsuki during the solar occultation experiment, the PSD slope confirms the dominance of coronal plasma fluctuations at small heliocentric offsets, where solar wind turbulence is strongly driven by both thermal and magnetic effects.

In contrast, the CH3PM S-band, VeRa S-band, and Akatsuki IPM-only X-band PSDs do not exhibit clear slopes. Additionally, for the CH3PM experiment, we see no effect in the signal due to the solar flare event that happened during that period. The absence of discernible power-law scaling indicates that plasma fluctuations in these datasets are weaker and relatively isotropic, dominated by low-amplitude, broadband noise arising from the quiet interplanetary medium far from the Sun and Moon. Nevertheless, mHz-level fluctuations are still measurable, particularly in CH3PM data, demonstrating that IPM can have sufficient perturbations to be detected in high-stability S-band two-way coherent signals, but not significant enough as compared to the contribution by the lunar environment as demonstrated using the CH2 data. These small-scale fluctuations are critical for quantifying the baseline contribution of the IPM to Doppler noise, which is essential for high-precision spacecraft tracking, deep-space navigation, and radio science experiments.

\begin{table*}[tbh]
\centering
\renewcommand{\arraystretch}{1.3}
\begin{tabular}{|c|c|c|c|c|c|c|c|}
\hline
\hline
\textbf{Date} &
\textbf{Probe} &
\textbf{Band} &
\textbf{Frequency} &
\textbf{Medium Probed} &
\textbf{Dst values} &
\textbf{Doppler FF} &
\textbf{Spectral Index} \\
 & & (S/X) & (MHz) & &(Daily avg) &(Hz) & ($\alpha$) \\
\hline
\hline
15 Dec 2008 & VeRa& S-band & 2300      & Venus-Earth IPM  &    -         & $\pm$2.5        & $0.01$ \\
\hline
08 Jun 2020 & Akatsuki & X-band & 8410      & Venus-Earth IPM  &    -         & $\pm$0.05        & $-0.28$ \\
\hline
23 Oct 2022 & Akatsuki & X-band & 8410      & Solar corona & 20    & $\pm$2           & $0.68$  \\
\hline
08 Nov 2022 & CH2      & S-band & 2041.598  & Lunar ionosphere& 50  & $\pm$0.075       & $2.18$  \\
\hline\
10 Nov 2025 & CH3PM    & S-band & 2028.78   & Earth-Moon IPM&  10 & $\pm$0.01 & $-0.68$ \\

\hline
\hline
\end{tabular}
\caption{Summary of spacecraft radio observations and plasma-induced Doppler frequency fluctuations.}
\label{tab:doppler_summary}
\end{table*}
Table \ref{tab:doppler_summary} shows the summary of the spacecraft radio observations and the observed Doppler frequency fluctuations (FF). The comparison between S-band signals and one-way X-band signals further highlights the sensitivity differences between frequencies and link configurations. Two-way S-band signals, as used in CH2 and CH3PM, effectively double the phase accumulation along the propagation path, enhancing the detectability of plasma-induced fluctuations even when amplitudes are low. This makes S-band particularly well-suited for studying the lunar ionosphere and the Moon-Earth interplanetary plasma. In contrast, one-way X-band signals, while probing longer interplanetary paths such as the Earth-Venus distance, accumulate less plasma-induced fluctuations. Although X-band signals are intrinsically less sensitive to plasma-induced effects due to their higher frequency, the Akatsuki observations considered here traverse long interplanetary paths, enabling measurable IPM-induced Doppler fluctuations. These observations are selected to avoid Venus’s atmosphere and the solar corona, thereby isolating the IPM contribution. Solar occultation measurements are used separately to probe plasma effects at smaller solar offsets.

The mHz-level fluctuations observed in CH3PM, VeRa, and Akatsuki IPM datasets indicate that the interplanetary medium, even in quiescent conditions, contains small-scale irregularities that can produce measurable Doppler shifts over large distances. These results provide a quantitative benchmark for the magnitude of IPM-induced noise in S-band and X-band radio signals, which is essential for designing future radio science experiments, calibrating deep-space navigation systems, and modeling the impact of plasma turbulence on spacecraft telemetry.

Combining these five datasets allows a comprehensive understanding of plasma-induced Doppler fluctuations across multiple environments, frequencies, and line-of-sight geometries. CH3PM, VeRa, and Akatsuki IPM observations provide direct measurements of the baseline interplanetary plasma effect on radio signals, which is crucial for interpreting S-band and X-band Doppler data free from near-planet or near-Sun contributions. CH2 observations provide a reference for near-lunar plasma and highlight the enhancement of fluctuations due to the lunar ionosphere and magnetotail. Akatsuki solar occultation data demonstrate the extreme plasma-induced Doppler effects in the solar corona, illustrating the upper bound of fluctuation amplitudes in a highly turbulent medium.

The turbulence characteristics presented here are derived from a limited number of representative radio-occultation experiments conducted under relatively quiet solar and geomagnetic conditions, enabling the plasma contributions from different propagation media to be isolated. Although these case studies demonstrate that the proposed spectral analysis can distinguish turbulence signatures associated with the terrestrial ionosphere, interplanetary medium, planetary ionospheres, and the lunar plasma environment, they do not capture the full statistical variability of these media. In particular, turbulence characteristics along the Earth--Venus propagation path are expected to depend on observing geometry, heliocentric distance, solar wind conditions, and solar activity. An additional limitation is illustrated by the analyzed solar flare event, for which no measurable change in the Doppler spectral slope was observed. This suggests that not all transient solar events produce sufficiently strong plasma perturbations along the propagation path to modify the observed turbulence characteristics, and that the spectral response depends on the disturbance intensity, propagation geometry, and the intervening plasma environment. Consequently, extending the present analysis to a substantially larger radio-occultation database spanning diverse observing geometries, solar flares, CMEs, solar wind conditions, and geomagnetic activity will be essential for quantifying these dependencies, establishing the robustness of the inferred turbulence parameters, and evaluating the sensitivity of the proposed spectral technique for space-weather diagnostics.

Overall, the PSD analysis demonstrates that plasma-induced Doppler fluctuations exhibit pronounced environment- and frequency-dependent characteristics. Two-way coherent S-band signals are highly sensitive to weak plasma turbulence, whereas X-band observations are better suited for probing stronger plasma environments over extended propagation paths. The observed spectral slopes and fluctuation amplitudes provide empirical constraints on models of interplanetary and coronal turbulence, support spacecraft mission planning, and contribute to the development of mitigation strategies for plasma-induced noise in high-precision radio science and navigation applications. More broadly, these results highlight the importance of multi-frequency observations acquired over diverse propagation geometries for comprehensively characterizing plasma turbulence throughout the inner solar system. A future statistical analysis of a larger radio-occultation database will further establish the general applicability of the proposed spectral approach and assess its potential as a diagnostic tool for monitoring plasma turbulence and identifying space-weather disturbances.

\section*{Acknowledgments}
K.A. would like to sincerely thank the Director of SPL and the Head of HRDD at VSSC, ISRO, for the opportunity to visit SPL as part of this project. The Prime Minister's Research Fellowship (PMRF) program, under the Ministry of Education, Government of India, awarded author K.A. a research scholarship (PMRF-2103356). Authors K.A. and A.D. acknowledge the use of facilities funded by the Department of Science and Technology, Government of India, through the DST-FIST grant no. SR/FST/PSII/2021/162 (C), awarded to DAASE at IIT Indore. 
We express our appreciation and gratitude to all members of the IDSN-18 and IDSN-32 ground station teams, the CH2/CH3 spacecraft operation teams, Ms. Pavithra R. Sindhe, the Flight Dynamics Team at ISTRAC, the Payload Planning teams, the Mission Directors, and the data center teams for their proactive support in conducting RO experiments in two-way mode. 
We would also like to thank the Akatsuki mission team and UDSC staff for their assistance in monitoring the Akatsuki radio signals. A special thanks to Himanshu Pandey and the ISSDC team for their kind help with data dissemination.


\end{document}